\documentclass[11pt]{article}

\usepackage{amsfonts}
\usepackage{latexsym}
\usepackage{amssymb}
\usepackage{amscd}
\usepackage{amsmath}
\newlength{\dinwidth}
\newlength{\dinmargin}
\def\be{\begin{equation}}
\def\ee{\end{equation}}
\def\ben{\begin{displaymath}}
\def\een{\end{displaymath}}
\def\baa{\begin{eqnarray}}
\def\eaa{\end{eqnarray}}

\def\ba{\begin{array}}
\def\ea{\end{array}}

\makeatletter
\@addtoreset{equation}{section}
\makeatother

\def\phi{\varphi}
\def\a{\alpha}
\def\b{\beta}

\def\de{{\delta}}
\def\e{\epsilon}
\def\l{\lambda}

\def\Acal{{\cal A}}
 
\def\Ucal{{\cal U}}
\def\Lhat{{\hat{{\cal L}}}}

\def\Pcal{{\cal P}}
\def\Ucal{{\cal U}}
\def\Ecal{{\cal E}}
\def\Wcal{{\cal W}}

\def\be{\begin{equation}}
\def\ee{\end{equation}}
\def\f{\frac}
\def\la{\label}

\def\CP1{{\mathbb C}P^1}
\def\la{\label}

\def\f{\frac}
\def\L{{\cal L}}
\def\p{\partial}
\def\tr{{\rm tr}}
\def\log{\ln}

\def\la{\label}

\def\f{\frac}
\def\p{\partial}
\def\res{{\rm res}}

\def\det{{\rm det}}

\def\B{{\bf B}}

\def\U{{V_1}}
\def\V{{V_2}}
\def\x{{f}}
\def\y{{g}}
\def\xp{{P}}
\def\yp{{Q}}
\def\ix{{\infty_f}}
\def\iy{{\infty_g}}

\def\dV{{d_2}}
\def\Y1{{Y^{(1)}}}

\def\B{{\bf B}}
\def\lp{{x}}

\def\blangle{{\Big\langle}}
\def\brangle{{\Big\rangle}}

\begin{document}

\begin{center}
\hskip9.0cm {\large Preprint SPhT-T04/010}
\vskip1.0cm
{\LARGE  $1/N^2$ correction to free energy in hermitian two-matrix model}
\vskip1.0cm
{\large B.Eynard$^1$, A.Kokotov$^2$, D.Korotkin$^2$}\\
\vskip0.5cm
$^1$ Service de Physique Th\'eorique, CEA/Saclay, Orme des Merisier
F-91191 Gif-sur-Yvette Cedex, France\\
$^2$ Concordia University, 7141 Sherbrooke West, Montreal H4B1R6,
Montreal, Quebec, Canada
\end{center}
\vskip1.0cm

{\bf Abstract.} Using the loop equations we find an explicit expression for genus 1 correction
in hermitian two-matrix model in terms of holomorphic objects
associated to spectral curve arising in large N limit. Our result
generalises
known expression for $F^1$ in hermitian one-matrix model. We discuss the
relationship
between $F^1$, Bergmann tau-function on
Hurwitz spaces, G-function of Frobenius manifolds and determinant of
Laplacian over spectral curve.

\vskip1.0cm

In this letter we derive an explicit formula for the $1/N^2$
correction to free energy $F$ of hermitian two-matrix model:
\be
e^{-N^2F}:= \int dM_1 dM_2 e^{-N\tr\{V_1(M_1)+V_2(M_2)-M_1 M_2\}}\;.
\la{part}
\ee
It is hard to overestimate the  interest to random matrix models in
modern physics and mathematics; 
we just mention their appearance in statistical physics,
condensed matter and 2d quantum gravity 
(see e.g.  \cite{rev}) . 
The expansion $F=\sum_{G=0}^{\infty} N^{-2G} F^{G}$ 
($N$  is the matrix size) in hermitian matrix models
is one of the cornerstones of the theory, due to its clear physical interpretation as topological expansion
of the functional integral, which appears in $N\to\infty$ limit; in statistical physics the term $F^{G}$ plays the
role of free energy for statistical physics model on genus $G$ Riemann  surface.
From the whole zoo if the random matrices one of the simplest is the hermitian 
one-matrix model  with partition function $e^{-N^2F}=\int dM e^{-N\tr
V(M)}$ (V is a polynomial), which can be used as testing ground for the methods applied in more general situations 
of two- and multi- matrix models.
The most rigorous way to compute the  $1/N^2$ expansion for both 
one-matrix and two-matrix models is based on the loop equations.
The loop equations follow from the  reparametrization invariance of matrix integrals; for one-matrix case
the loop equations were used to compute $F^1$ (see \cite{onematr}).
Later  the loop equations were written down for the case of two-matrix model \cite{Staudacher,Eynard}
 and  $F^1$  was found
for the case when the spectral curve has genus zero and one \cite{Eynard}; 
for arbitrary genus of spectral curve of two-matrix model
only the leading term $F^0$ is known (see  \cite{Marco}).

Let us write down polynomials $\U$ and $\V$ in the form
$\U(x)=\sum_{k=1}^{d_1+1} \f{u_k}{k} x^k$ and 
$\V(y)=\sum_{k=1}^{d_2+1} \f{v_k}{k} y^k$. It is sometimes convenient to think of $\U$ and $\V$ 
as infinite formal power expansions: $\U(x)=\sum_{k=1}^{\infty} \f{u_k}{k} x^k$, $\V(y)=\sum_{k=1}^{\infty} \f{v_k}{k} y^k$,
where coefficients $u_k$ vanish for $k\geq d_1+2$, and $v_k$ vanish for $k\geq d_2+2$. According to this point of view
the operators of differentiation with respect
to coefficients of $\U$ and $\V$ have the following meaning (see \cite{Marco}):
\be
\f{\de}{\de\U(x)}\Big|_x:=\left\{\sum_{k=1}^{\infty}x^{-k-1} k\p_{u_k}\right\}
\Big|_{u_k=0\;, k\geq d_1+2}\;,\;\;\;\hskip0.8cm
\f{\de}{\de\V(y)}\Big|_y:=\left\{\sum_{k=1}^{\infty}y^{-k-1} k\p_{v_k}\right\}\Big|_{v_k=0\;, k\geq d_1+2}\;.
\la{formal}
\ee
As it was discussed in detail in \cite{Marco}, (\ref{formal})
is a formal notation which makes sense only 
order by order in the infinite power series expansion; it allows to write an 
infinite number of equations at once.
% and significantly shortens the presentation.
%In particular, according to these notations,
%\be
%\f{\de \U(x)}{\de \U(\xt)}=\f{1}{\xt-x}\;,\hskip0.6cm 
%\f{\de \U'(x)}{\de \U(\xt)}=\f{1}{(\xt-x)^2}\;.
%\la{deltaf}
%\ee
Consider the resolvents
$\Wcal(x)=\f{1}{N}\blangle \tr\f{1}{x-M_1}\brangle$ and
$\tilde{\Wcal}(y)=\f{1}{N}\blangle \tr\f{1}{y-M_2}\brangle$
The free energy of two-matrix model (\ref{part}) satisfies the following equations with respect to
coefficients of polynomials $\U$ and $\V$:
\be
\f{\delta F}{\delta \U(x)} =\Wcal(x)\;,\hskip0.7cm
\f{\delta F}{\delta \V(y)} =\tilde{\Wcal}(y)\;.
\la{varint}
\ee
The equations (\ref{varint}) 
were solved in \cite{Marco} in the zeroth order assuming the finite-gap structure
of distribution of eigenvalues of $M_1$ (and, {\it a posteriori}, also of $M_2$) as $N\to\infty$.
Here we find the next coefficient $F^1$,
using the loop equations.  
The spectral curve ${{\cal L}}$ is defined by the following equation, which arises in the zeroth order approximation:
\be
\Ecal^{0}(x,y):= (\U'(x)-y)(\V'(y)-x) -\Pcal^{0}(x,y)+1=0
\la{Lint}
\ee
 where the polynomial of two variables 
$\Pcal^{0}(x,y)$ is the zeroth order term in $1/N^2$ expansion of the polynomial  
\be
\Pcal(x,y):=\f{1}{N}\blangle \tr\f{\U'(x)-\U'(M_1)}{x-M_1}\f{\V'(y)-\V'(M_2)}{y-M_2}\brangle\;;
\la{Pxy}
\ee
the point $P\in{{\cal L}} $ of the curve is the pair of complex numbers $(x,y)$ satisfying (\ref{Lint})
(on the ``physical" sheet the equation of spectral curve (\ref{Lint}) defines an implicit
function $y(x)$, which gives the zeroth order approximation to $\V'(x)-\Wcal(x)$).
The spectral curve (\ref{Lint}) comes together with two meromorphic functions
$\x(P)=x$ and $\y(P)=y$, which project it down to $x$ and $y$-planes, respectively.
These functions have poles only at two points of ${{\cal L}}$, called $\ix$ and $\iy$: at $\ix$ function $\x(P)$ has
simple pole, and function $\y(P)$ - pole of order $d_1$ with singular part equal to $\U'(\x(P))$. At
$\iy$ the function $\y(P)$ has simple pole, and function $\x(P)$ - pole  of order $d_2$ with singular part
equal to $\V'(\y(P))$. In addition, in our normalization of partition function (\ref{part}) we 
have the asymptotics
$\Wcal(x)\sim_{x\to\infty} 1/x+\dots$ and $\tilde{\Wcal}(y)\sim_{y\to\infty} 1/y+\dots$, which imply \cite{Marco}
${\rm Res}_{\infty_f} g df = 1$ and ${\rm Res}_{\infty_g} f dg=1$, respectively.
 Therefore, one gets  the moduli space ${\cal M}$ of triples $({{\cal L}},\x,\y)$, where functions $\x$ and $\y$ have this pole structure. The natural coordinates on this moduli space are coefficients of polynomials $\U$ and $\V$ and $g$ numbers,
called ``filling fractions" $\e_\a=\f{1}{2\pi i}\oint_{a_\a}\y d\x$, where $(a_\a,b_\a)$ is some basis of canonical cycles on ${{\cal L}}$.  
The additional constraints which should be imposed {\it a posteriori}
to make the  ``filling fractions''  dependent on coefficients of 
polynomials $V_1$ and $V_2$ are (according to one-matrix model
experience, these conditions correspond to non-tunneling between
different intervals of eigenvalues support):
$\oint_{b_a}\y d\x=0$.
Denote the zeros of differential $d\x$ by $\xp_1,\dots,\xp_{m_1}$ ($m_1=d_2+2g+1$) (these points play the 
role of ramification points if we realize $\L$ as branched covering by function $\x(P)$); their projections on 
$x$-plane are the branch points, which we denote we denote by $\l_j:=\x(\xp_j)$ .
  The zeros of the differential $d\y$ (the ramification points if we consider ${{\cal L}}$ 
as covering defined by function $\y(P)$) we denote by 
$\yp_1,\dots,\yp_{m_2}$  ($m_2=d_1+2g+1$); 
there projections on $y$-plane (the branch points) we denote by $\mu_j:=\y(\yp_j)$.
We shall assume hat our potentials $\U$ and $\V$ are generic i.e. all zeros of differentials $d\x$ and $d\y$ are simple, and none of the zeros of $d\x$ coincides with a zero of $d\y$.
If is well-known (see for instance \cite{Marco}) how to express all standard algebro-geometrical objects on ${{\cal L}}$ in terms of the previous  data. In particular, the Bergmann bidifferential $B(P,Q)=d_P d_Q\log E(P,Q)$ ($E(P,Q)$ is the prime-form)  can be
represented as follows: 
%\begin{lemma}
\be
 B(P,Q)=\f{\de \y(P)}{\de \U(\x(Q))}\Big|_{\x(Q)} d\x(P) d\x(Q)
\la{Bx}
\ee
 The Bergmann bidifferential has the following 
behaviour near diagonal $P\to Q$:
$B(P,Q)=\{(z(P)-z(Q))^{-2}+\- \f{1}{6}S_B(P)+\- o(1)\} d z(P) d z(Q)$,
where $z(P)$ is some local coordinate; $S_B(P)$ is the Bergmann projective connection .
Consider also the  four-differential $D(P,Q)=d_P d_Q^3\log E(P,Q)$, which has on the diagonal the pole of 4th degree:
$D(P,Q)=\{ 6(z(P)-z(Q))^{-4}+ O(1)\} d z(P) (d z(Q))^3$.
From $B(P,Q)$ and $D(P,Q)$ it is easy to construct meromorphic normalized (all $a$-periods vanish) $1$-forms on ${{\cal L}}$ 
with single pole; in particular, if the pole coincides with ramification point $\xp_k$, the natural local parameter
near $\xp_k$ is $\lp_k(P)=\sqrt{\x(P)-\lambda_k}$; then
$B(P,\xp_k):=\f{B(P,Q)}{d\lp_k(Q)}\Big|_{Q=P_k}$ and
$D(P,\xp_k):=\f{D(P,Q)}{(d\lp_k(Q))^3}\Big|_{Q=P_k}$
are meromorphic normalized 1-forms on $\L$ with single pole at $\xp_k$ and the following singular parts:
\be
B(P,\xp_k)=\left\{\f{1}{[\lp_k(P)]^2}+ \f{1}{6}S_B(P_k)+ o(1)\right\} d\lp_k(P)\;;\hskip0.7cm
D(P,\xp_k)=\left\{\f{6}{[\lp_k(P)]^4}+ O(1)\right\} d\lp_k(P)
\la{BDk}
\ee
as $P\to \xp_k$, where $S_B(P_k)$  is the Bergmann projective connection computed at the branch point $P_k$
with respect to the local parameter $\lp_k(P)$.

Equations (\ref{varint}) in order $1/N^2$ look as follows (we write only equations with respect to $\U$):
\be
\f{\de F^{1}}{\de\U(\x(P))}=-Y^{1}(P)
\la{F1Yint}
\ee
where the $Y^{1}$ is the (taken with minus sign) $1/N^2$ contribution
to the resolvent 
${{\cal W}}$. The function $Y^{1}$
can be computed using the
loop equations \cite{Eynard} and the ``normalization conditions''
\be
\oint_{a_\a}Y^{1}(P)d\x(P)=0
\la{assum}
\ee
over all basic $a$-cycles (these conditions mean that the ``filling fractions''
do not have the $1/N^2$ correction).
We introduce also the
polynomial 
\be
\Ecal(x,y):=(\U(x)-y)(\V(y)-x)-\Pcal(x,y)+1\;,
\la{Exy}
\ee
the function  $\Ucal(x,y)$, which is a polynomial in $y$ and rational function in $x$:
\be
\Ucal(x,y):=\f{1}{N}\blangle \tr\f{1}{x-M_1}\f{\V'(y)-\V'(M_2)}{y-M_2}\brangle\;,
\la{Uxy}
\ee
and rational function $\Ucal(x,y,z)$:
\be
\Ucal(x,y,z):=\f{\de \Ucal(x,y)}{\de \U(z)}= 
\blangle \tr\f{1}{x-M_1}\f{\V'(y)-\V'(M_2)}{y-M_2}\tr\f{1}{z-M_1}\brangle
-N^2 \Ucal(x,y) \Wcal (z)\;.
\la{Uxyxt}
\ee
Then the  loop equation looks as follows:  
\be
\Ucal(x,y)=x-\V'(y)+\f{\Ecal(x,y)}{y-Y(x)}-\f{1}{N^2}\f{\Ucal(x,y,x)}{y-Y(x)}\;,
\la{loop1}
\ee
where $Y(x):=\U'(x)-\Wcal(x)$; it arises as a corollary of reparametrization invariance of the matrix integral
(\ref{part}) \cite{Eynard}.
To use the loop equation  effectively we need to consider the $1/N^2$ expansion of all of their ingredients.
In this way we get the following expression for $Y^1$:
\be
Y^1(P)d\x(P)=\f{\Pcal^1(\x(P),\y(P))d\x(P)}{\Ecal^0_y(\x(P),\y(P))}+\sum_{Q\neq P\;:\;\x(Q)=\x(P)}\f{B(P,Q)}{d\x(Q)}
\f{1}{\y(P)-\y(Q)}\;,
\la{Y111}
\ee
where $\Ecal^0_y(x,y)$ means partial derivative with respect to the second argument. 
All ingredients of (\ref{Y111}) arise already in the leading term,
except $\Pcal^1$. However, from (\ref{Pxy})  we see  that $\Pcal(x,y)$
is a polynomial  of degree $d_1-1$ with respect to $x$ and $d_2-1$
with respect to $y$; moreover, the coefficient in front of $x^{d_1-1}
y^{d_2-1}$ does not have $1/N^2$ correction. Thus we can conclude that
the one-form $Y^1(P)d\x(P)$
is non-singular on the spectral curve outside of the branch points
$P_m$ (where it has poles of order 4); moreover, the first term in (\ref{Y111}) is non-singular on
${{\cal L}}$ (the  first order zeros of $\Ecal^0_y$ at the branch points are
cancelled by first order zeros of $d\x(P)$ at these points). 
The form of singular parts at $P_m$ allows to determine $Y^1(P)d\x(P)$
completely in terms of differentials
$B(P,P_k)$ and $D(P,P_k)$ if we take into account the absence of
$1/N^2$ correction to the ``filling fractions'' (\ref{assum}); the result looks as follows:
\be
\Y1(P)d\x(P) =\sum_{k=1}^{m_1}\left\{-\f{1}{96\y'(\xp_k)}D(P,\xp_k) + \left[\f{\y'''(\xp_k)}{96\y'^2(\xp_k)} -\f{S_B(\xp_k)}{24\y'(\xp_k)}\right] B(P,\xp_k)\right\}
\la{Y1int}
\ee
Then the solution of (\ref{F1Yint}), (\ref{Y1int}), which is symmetric with respect to the projection change
(and, therefore,   satisfies also  equations (\ref{varint}) with respect to $\V$), looks as follows:
\be
F^{1}=\f{1}{24}\log\left\{\tau^{12}_{\x} (v_{\dV+1})^{1-\f{1}{\dV}}\prod_{k=1}^{m_1} d\y(\xp_k) \right\}+
\f{d_2+3}{24}\log d_2
\la{F1int}
\ee
where $\tau_\x$ is the so-called Bergmann tau-function on Hurwitz
 space \cite{KK}, which satisfies the following system 
of equations with respect to the branch points $\lambda_k$:
\be
\f{\p}{\p \lambda_k}\log\tau_{\x} = -\f{1}{12} S_B(P_k)\;;
\la{bergint}
\ee
In derivation of (\ref{Y1int}) we have used the following variational formulas, which can be easily proved in analogy
to Rauch variational formulas:
\be
-\f{\de\lambda_k}{\de \U(\x(P))} \y'(P_k) d\x(P)= B(P,P_k)\;,
\la{zerothor}
\ee
\be
\f{\de\{\y'(\xp_k)\}}{\de \U(\x(P))}\Big|_{\x(P)} d\x(P)=\f{1}{4}\left\{D(P,\xp_k)-\f{\y'''(\xp_k)}{\y'(\xp_k)} B(P,\xp_k)\right\}
\la{yV1}
\ee

The Bergmann tau-function (\ref{bergint}) appears in many important problems: it coincides with isomonodromic
tau-function of Hurwitz Frobenius manifolds \cite{Dub92}, and gives the main contribution to $G$-function
(solution of Getzler equation)
of these Frobenius manifolds; it gives the most non-trivial term in isomonodromic tau-function of Riemann-Hilbert problem with quasi-permutation monodromies. Finally, its modulus square essentially coincides with determinants of Laplace
operator in metrics with conical singularities over Riemann surfaces \cite{KK}. 
The solution of the system (\ref{bergint}) looks as follows \cite{KK}. 
Define the divisor $(d\x)=-2\ix-(d_2+1)\iy+\sum_{k=1}^{m_2} P_k:=\sum_{k=1}^{m_2+2}r_k D_k$.
Choose some initial point 
$P\in\Lhat$ and introduce corresponding vector of Riemann constants $K^P$ and Abel map $\Acal_\a(Q)=\int_P^Q w_\a$
($w_\a$ form the basis of normalized holomorphic 1-forms on ${{\cal L}}$).
Since some points of divisor $(d\x)$ have multiplicity 1, we can always choose the fundamental cell $\Lhat$
of the universal covering of ${{\cal L}}$  in such a way 
that $\Acal((d\x))=-2K^P$ (for an arbitrary choice of fundamental domain these two vectors coincide only up to
an integer combination of periods of holomorphic differentials), where the Abel map is computed along the path which does not intersect the boundary of $\Lhat$.
The main ingredient of the Bergmann tau-function is the following
holomorphic multivalued $(1-g)g/2$-differential ${{\cal C}}(P)$ (the higher genus analog of Dedekind eta-function) on ${{\cal L}}$:
\be
{{\cal C}}(P):=\f{1}{W(P)}{\sum_{\a_1,\dots,\a_g=1}^g 
\frac{\partial^g\Theta(K^P)}{\p z_{\a_1}\dots \p z_{\a_g}} w_{\a_1}(P)\dots w_{\a_g}(P)}\;.
\ee
where
$
W(P):= {\rm \det}_{1\leq \a, \b\leq g}||w_\b^{(\a-1)}(P)||
$
denotes the Wronskian determinant of holomorphic differentials
 at point $P$; $K^P$ is the vector of Riemann constants with basepoint $P$. 
Introduce the quantity ${\cal Q}$  defined by
\begin{equation}
{\cal Q} = [d\x(P)]^{\f{g-1}{2}} {{\cal C}}(P)
\prod_{k=1}^{m+N}[ E (P,D_k)]^{\f{(1-g)r_k}{2}}\;;
\la{Fdef1}
\end{equation}
which is independent of the point $P\in {{\cal L}}$. 
Then the Bergmann tau-function (\ref{bergint}) of Hurwitz space  is given by the following expression \cite{KK}:
\begin{equation}
\tau_{\x} = {{\cal Q}}^{2/3}  \prod_{k,l=1\;\;k< l}^{m+n} [E(D_k, D_l)]^{\frac{r_k r_l}{6}} \;; 
\la{taui}
\end{equation}
together with (\ref{F1int}) this gives the answer for $1/N^2$ correction in two-matrix model.
If $\tau_{\x}$ and $\tau_{\y}$ are Bergmann tau-functions (\ref{bergint}) corresponding to 
divisors $(d\x)$ and $(d\y)$,
respectively, then
\be
\left(\f{\tau_{\x}}{\tau_{\y}}\right)^{12}=C\f{(u_{d_1+1})^{1-\f{1}{d_1}}}{(v_{d_2+1})^{1-\f{1}{d_2}}}
\f{\prod_{k}d\x(\yp_k)}{\prod_k d\y(\xp_k)}
\la{ttau}\ee
where $C={d_1^{d_1+3}}/{d_2^{d_2+3}}$
 is a constant independent of moduli parameters.
Using the transformation (\ref{ttau}) of the Bergmann tau-function under projection change,
we find that the solution expression (\ref{F1int}) for $F^1$ satisfies also the necessary equations with respect  to $\V$. This could be considered as a confirmation of consistency of our computation.
Derivatives of function $F^1$  (\ref{F1int}) with respect to the filling fractions look as follows:
\be
\f{\p F^1}{\p \e_\a}=-\oint_{b_\a}Y^1(P)d\x(P)\;;
\ee
these equations are $1/N^2$ counterparts of Seiberg-Witten type equations
for $F^0$  (see for example \cite{Eynard,Marco,KazMar}).

{\it One-matrix model.} If potential $\V$ is quadratic, integration with respect to $M_2$ in (\ref{part}) can be performed explicitly, and
the free energy (\ref{F1int}) gives rise to the free energy of one-matrix model. The spectral curve ${{\cal L}}$ in this
case  becomes hyperelliptic, and the formula (\ref{F1int}) gives, using the expression for $\tau_\x$ obtained in 
\cite{KitKor}:
\be
F^1=\f{1}{24}\log\left\{\Delta^3\,(\det{\bf A})^{12}  \prod_{k=1}^{2g+2} \y'(\lambda_k)\right\}
\la{F1hypint}
\ee
where $\lambda_k$, $k=1,\dots,2g+2$ are branch points of ${{\cal L}}$; $\Delta$ is their Wronskian determinant; ${\bf A}$ 
is the matrix of $a$-periods of non-normalized holomorphic differentials on ${{\cal L}}$; this agrees 
with previous results \cite{onematr}.

{\it $F^1$, isomonodromic tau-function  and $G$-function of Frobenius manifolds.}
The genus 1 correction to free energy in topological field theories is given by
so-called $G$-function, which for an arbitrary $m$-dimensional 
Frobenius manifold related to Hurwitz space looks as follows \cite{Dub92,DZ,KK}:
$\exp\{G\}={\tau_f^{-1/2}}{\prod_{k=1}^m \{\res|_{P_k}\f{\varphi^2}{d\x}\}^{-1/48}}$,
where $\tau_f$ is the Bergmann tau-function, $\varphi$ is an ``admissible" one-form on underlying
Riemann surface.
If, trying to build an analogy with our formula (\ref{F1int}) for $F^1$, we formally choose $\phi(P)=d\y(P)$,
the formula for G-function coincides with (\ref{F1int}) up to 
small details like sign, additive constant, and the highest coefficient of polynomial $\V$.
However, the differential $d\y$ is not admissible, and, therefore, does not really correspond to
any Frobenius manifold; therefore, the true origin of his analogy is still unclear

{\it $F^1$ and determinant of Laplace operator.}
Existence of close relationship between $F^1$ and determinant of Laplace operator was suggested
by several authors (see e.g. \cite{DV} for  hermitial one-matrix model, \cite{Eynard}
for hermitian two-matrix model and, finally, \cite{Zabr} for normal two-matrix model 
with simply-connected support of eigenvalues). After an  appropriate regularization 
the (formal) determinant of Laplace operator $\Delta^\x$ over ${{\cal
L}}$ in the singular metric $|d\x(P)|^2$ is given by the expression \cite{Sonoda}
$
{\det \Delta^\x} = C \,{{\cal A}\,\{\det \Im \B}\}|\tau_\x|^2\;,
$
where ${{\cal A}}$ is a regularized area of $\L$, $\B$ is the matrix of $b$-periods of $\L$, $C$ is a constant.
In the ``physical" case of real coefficients of $\U$ and $\V$ and real filling fractions
 the empirical expression  for 
$\log\{\det \Delta^\x\}$ differs from  our expression (\ref{F1int}) by several explicit terms.
Therefore, the  relationship between hermitian and normal two-matrix models
 on the level of $F^1$ seems to be  not as
straightforward as on the level of functions $F^0$ \cite{Marco,Zabr}. 

{\bf Acknowledgements} We thank M.Bertola, L.Chekhov, B.Dubrovin,
T.Grava, V.Kazakov, I.Kostov, M.Staudacher and S.Theisen for important discussions.
The work of BE was partially supported by the EC ITH Network HPRN-CT-1999-000161.
The work of DK was partially supported by  NSERC, NATEQ and Humboldt foundation. 
AK and DK thank Max-Planck Institute for Mathematics in Bonn for support and nice working conditions.
DK thanks also  SISSA and CEA  for support and hospitality.

\end{document}